\documentclass[a4paper,11pt]{article}
\pdfoutput=1 

\usepackage{jinstpub} 

\usepackage{lineno}
\usepackage{gensymb}

\nolinenumbers

\title{\boldmath Real-time Multi-Messenger Analysis Framework for KM3NeT}

\author[a]{W. Assal,}
\author[a,1]{D. Dornic,}
\author[a,1]{F. Huang,\note{Corresponding authors.}}
\author[a]{E. Le Guirriec,}
\author[a,b]{M. Lincetto,}
\author[a,c]{and G. Vannoye}

\affiliation[a]{Aix Marseille Univ, CNRS/IN2P3, CPPM, Marseille, France }
\affiliation[b]{Ruhr University Bochum, Germany}
\affiliation[c]{Ecole Normale Supérieure de Lyon, 15 parvis René Descartes BP 7000 69342 Lyon Cedex 07 France}

\emailAdd{dornic@cppm.in2p3.fr}
\emailAdd{feifei.huang@cppm.in2p3.fr}

\abstract{KM3NeT is a multi-purpose cubic-kilometer neutrino observatory under construction in the Mediterranean Sea. It consists of ORCA and ARCA (for Oscillation and Astroparticle Research with Cosmics in the Abyss, respectively), currently both have a few detection lines in operation. Although having different primary goals, both detectors can be used for neutrino astronomy over a wide energy range, from a few tens of MeVs to a few tens of PeV. In view of the growing field of time-domain astronomy, it is crucial to be able to identify neutrino candidates in real-time. This online neutrino sample will allow triggered neutrino alerts that will be sent to the astronomy community and to look for time/space coincidences around external electromagnetic and multi-messenger triggers. These real-time searches can significantly increase the discovery potential of transient cosmic accelerators and refine the pointing directions in the case of poorly localized triggers, such as gravitational waves. In the field of core-collapse supernovae (CCSN), the detection of the MeV-scale CCSN neutrinos is crucial as an early warning of the electromagnetic follow-up. KM3NeT’s digital optical modules act as good detectors for these supernovae neutrinos. This proceeding presents the status of KM3NeT's real-time multi-messenger activities, including online event reconstruction, event classification and selection, alert distribution, and supernova monitoring.} 


\arxivnumber{2107.13908} 

\collaboration[c]{on behalf of the KM3NeT collaboration}

\proceeding{VLVnT 2021 - Very Large Volume Neutrino Telescope Workshop\\
18-21 May 2021 \\
Valencia}

\begin{document}
\maketitle
\flushbottom

\section{KM3NeT Detectors and Science}
\label{sec:intro}
KM3NeT~\cite{loi}, the Cubic Kilometre Neutrino Telescope being built in the Mediterranean Sea, will detect neutrinos in a wide energy range. More details about the KM3NeT detectors can be found in the
proceedings of ref.~\cite{rosa}. This proceeding describes the status of KM3NeT's real-time multi-messenger analysis framework.

\section{Real-time Analysis Framework}
The KM3NeT real-time alert analysis framework consists of two pipelines: 1) the MeV CCSN monitoring pipeline; 2) the GeV-PeV neutrino alert pipeline. The main goals are: 1) CCSN monitoring; 2) receive external electromagnetic (EM), gravitational waves or neutrino alerts and search for correlated neutrinos; 3) send all-flavor all-sky neutrino alerts (e.g. multiplets, HE) to external observatories for follow-up. 

\begin{figure}[htbp]
\centering
\includegraphics[width=0.57\textwidth]{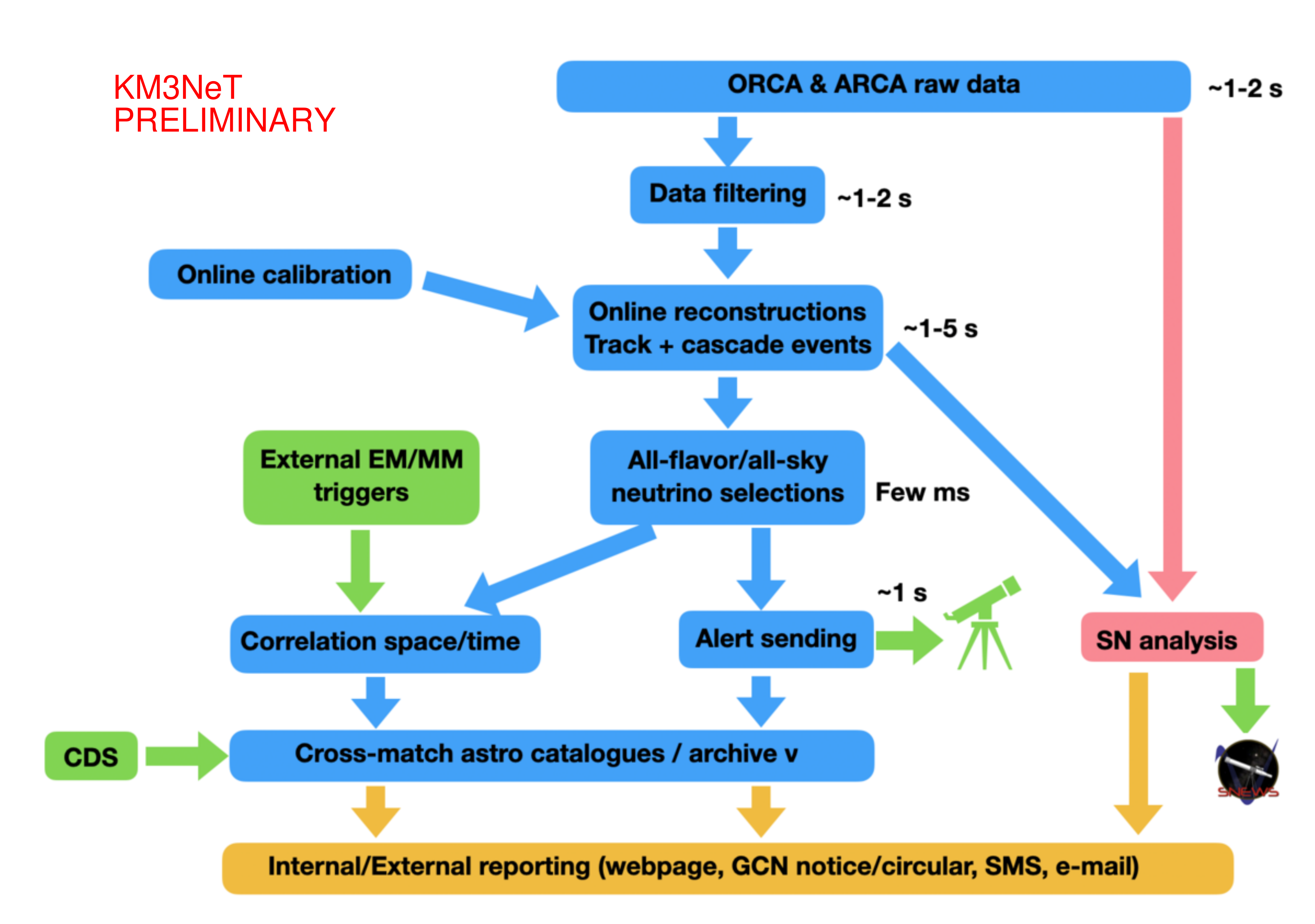}
\includegraphics[width=0.41\textwidth]{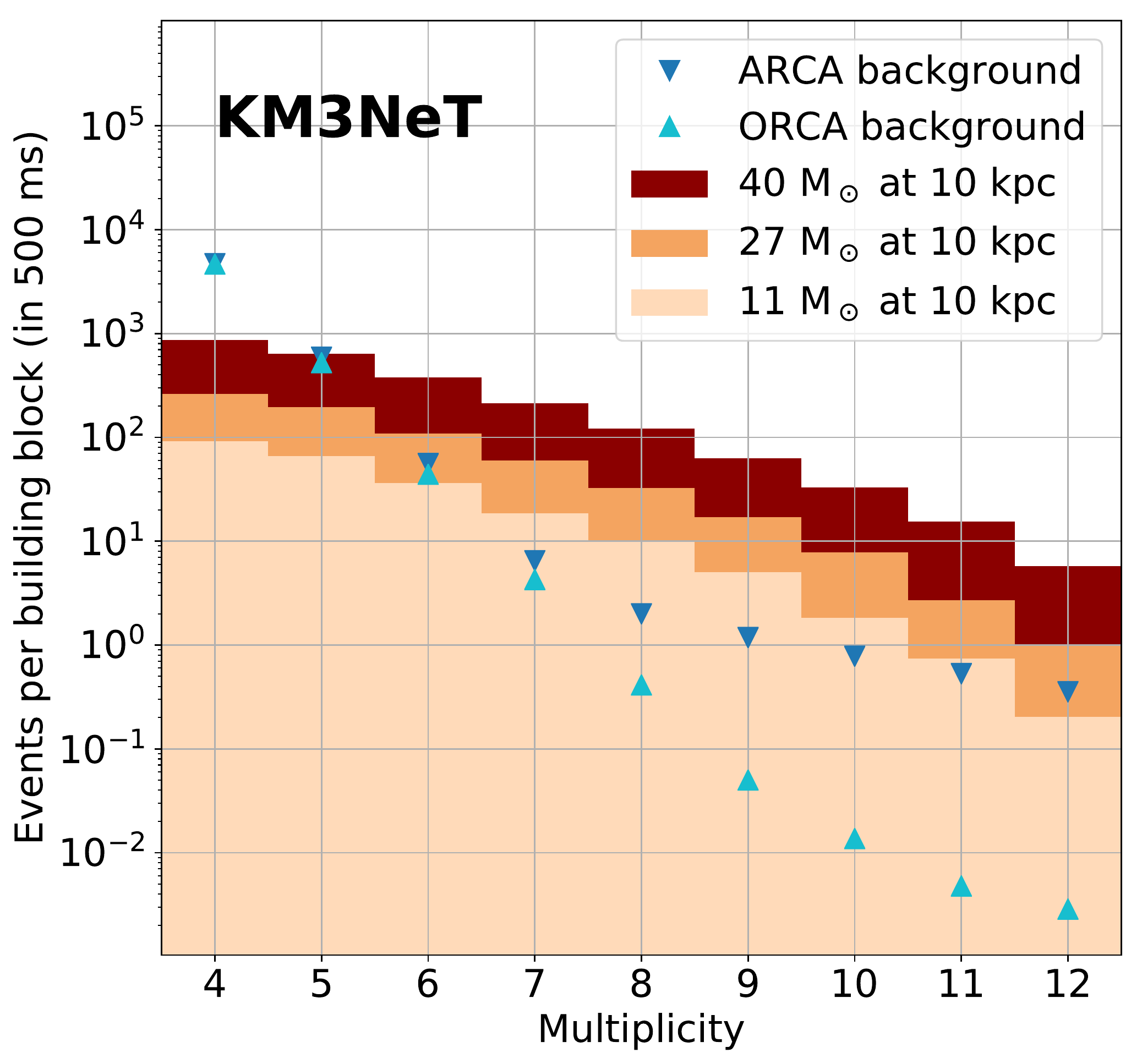}
\caption{Left: Overview of the KM3NeT real-time analysis framework. The green boxes show the external EM and multi-messenger (MM) alerts and the catalogs of the astronomical objects outside the solar system from CDS (Astronomical Data Center). 
The yellow box shows the internal or external reporting tools, mainly through GCN\protect\footnotemark. Right: Supernova events expected from 3 simulated progenitors at ORCA and ARCA as a function of multiplicity values compared with background rates\cite{ccsn}.}
\label{fig:2}
\end{figure}

Figure~\ref{fig:2} (left) shows the data flow and main components of the framework. The blue and green boxes and arrows are for the GeV-PeV neutrino pipeline and the red box and arrow for the MeV CCSN pipeline. Speed is essential for issuing multi-messenger alerts; our framework works (nearly) real-time with a response time on the order of 10~s.

\subsection{MeV CCSN Monitoring Pipeline}

The detection of supernova neutrinos is crucial for CCSN observation because they arrive hours before the photons and thus can act as an early warning. 
The CCSN pipeline monitors the coincidences of PMT hits inside each DOM (multiplicity). Figure~\ref{fig:2} (right) (taken from ref.~\cite{ccsn}) shows the signal vs. background rates expected from three simulated progenitors as a function of multiplicity. This pipeline is operational and sending alerts with false alarm rate less than 1/week to SNEWS~\cite{snews, snews2}. The alert generation latency is less than 20~s.

\subsection{GeV-PeV Neutrino Processing Pipeline}

In the neutrino processing pipeline, physics events are first triggered, then reconstructed and classified. This is done separately for ORCA and ARCA at their shore stations. ORCA and ARCA share the same processing structure and the same track reconstruction algorithms (but they have different shower reconstructions). After the processing, their data streams are combined for further analysis and alert sending (if criteria are satisfied). 
\footnotetext{The Gamma-ray Coordinates Network, https://gcn.gsfc.nasa.gov}

\subsubsection{Online Reconstruction}
\label{section:reco}

\begin{figure}[htbp]
\centering
\includegraphics[width=.49\textwidth]{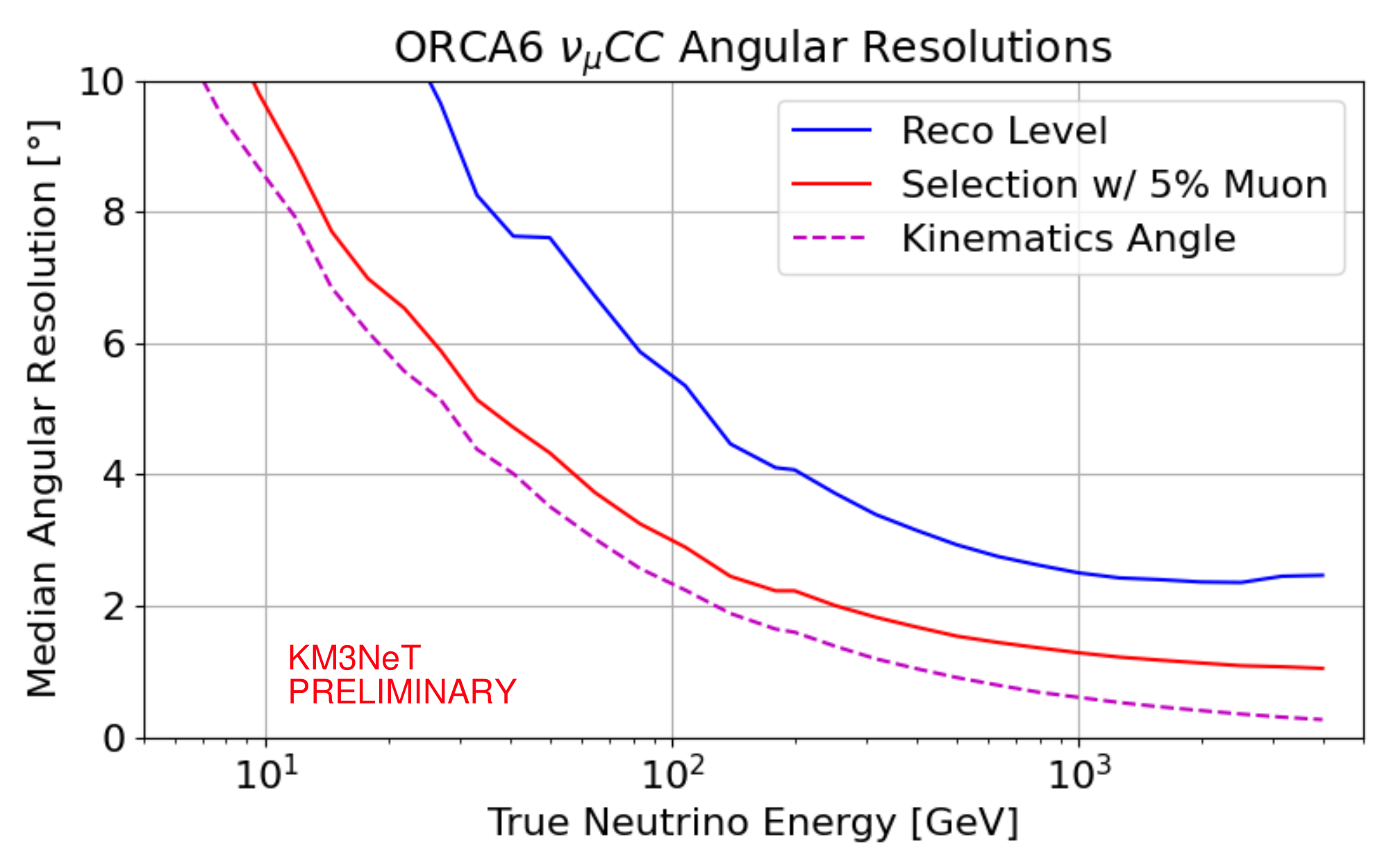}
\includegraphics[width=.49\textwidth]{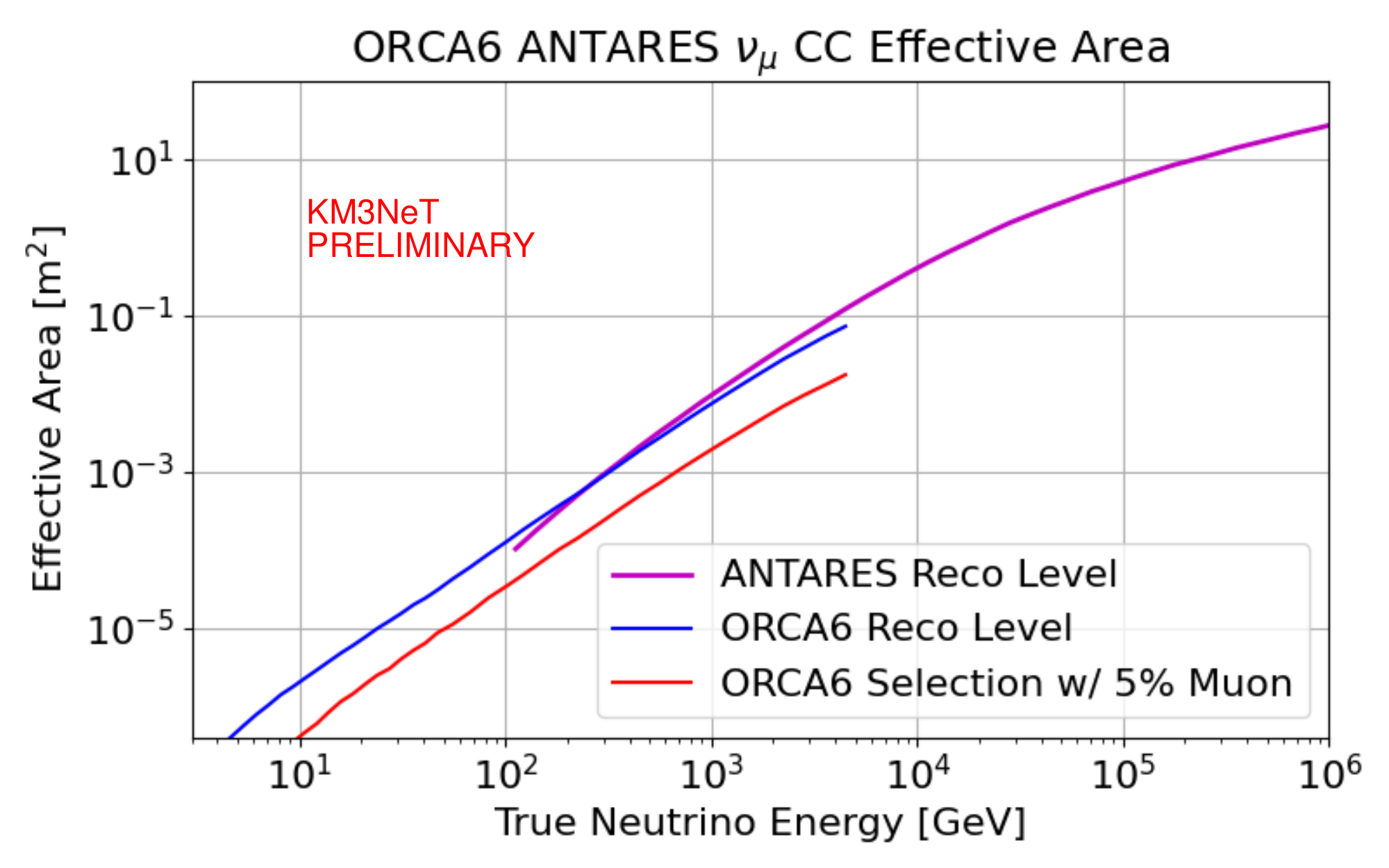}
\qquad
\caption{Left: ORCA6 median angular resolutions at the reconstruction level and at the preliminary selection with a 5\% muon contamination, and the kinematics angle. Right: Effective area comparisons at ANTARES and ORCA6 reconstruction level and the ORCA6 selection, the fraction of events selected is uniform over the energy range.}
\label{fig:3}
\end{figure}

KM3NeT event reconstruction consists of track reconstruction~\cite{reco} and shower reconstruction~\cite{shower_alba}. Online reconstruction uses the same algorithms as offline, although it does not run the complete reconstruction chain for every event. Optimizations are underway, for example, the shower reconstruction may be skipped for events that are clearly atmospheric muons. For ORCA6, it takes on average 0.3~s per event for the track reconstruction and 1~s for shower. Events with more hits or more DUs will take longer to reconstruct, but in general stays on the order of 10~s. 
The muon track of track-like events gives a good directional resolution. Figure~\ref{fig:3} (left) shows the median angular resolutions of ORCA6 $\nu_\mu$ CC events at reconstruction level (blue) and at a preliminary selection based on the track classification with a 5\% muon contamination (red), which achieves 8$\degree$ at 20 GeV, less than 2$\degree$ at above 300 GeV, and around 1$\degree$ at the TeV scale. The angular resolution is close to the true $\nu-\mu$ kinematics angle (dashed magenta), though still limited by the detector size and will improve as more DUs get implemented. 

\subsubsection{Online Track Classification}
\label{section:sel}

The online classification is able to extract neutrinos out of the enormous muon background and it works fast (0.01~s). The online track classifier was trained with a gradient boosting decision tree algorithm~\cite{xgboost} with MC $\nu_\mu$ CC events as signals and MC atmospheric muons as background. The classifier assigns each event a score in the range [0, 1], indicating the probability of it being a signal neutrino. The training features are 14 parameters that provide good signal background separation, for example the reconstructed track direction, the ratio of the track reconstruction's likelihood and number of degrees of freedom, the reconstructed neutrino interaction vertex position, the sum of Time Over Threshold (proxy for the charge observed by the triggered DOMs), and other geometry, charge and time-based parameters. Training features are selected by removing the less contributing ones. Then the classifier's hyperparameters that control the learning of the classifier are tuned, such as learning rate, number of boosting rounds, maximum tree depth, etc. After the classifier tuning and training, a test set (different from the training set to ensure no bias) is used to evaluate the classifier's performance. Figure~\ref{fig:3} shows the resolution and effective area with a preliminary selection that gives a 5$\%$ muon contamination.

\subsubsection{Preliminary Selection based on Online Classifier}
\begin{figure}[htbp]
\centering
\includegraphics[width=.46\textwidth,trim=12 0 35 0,clip]{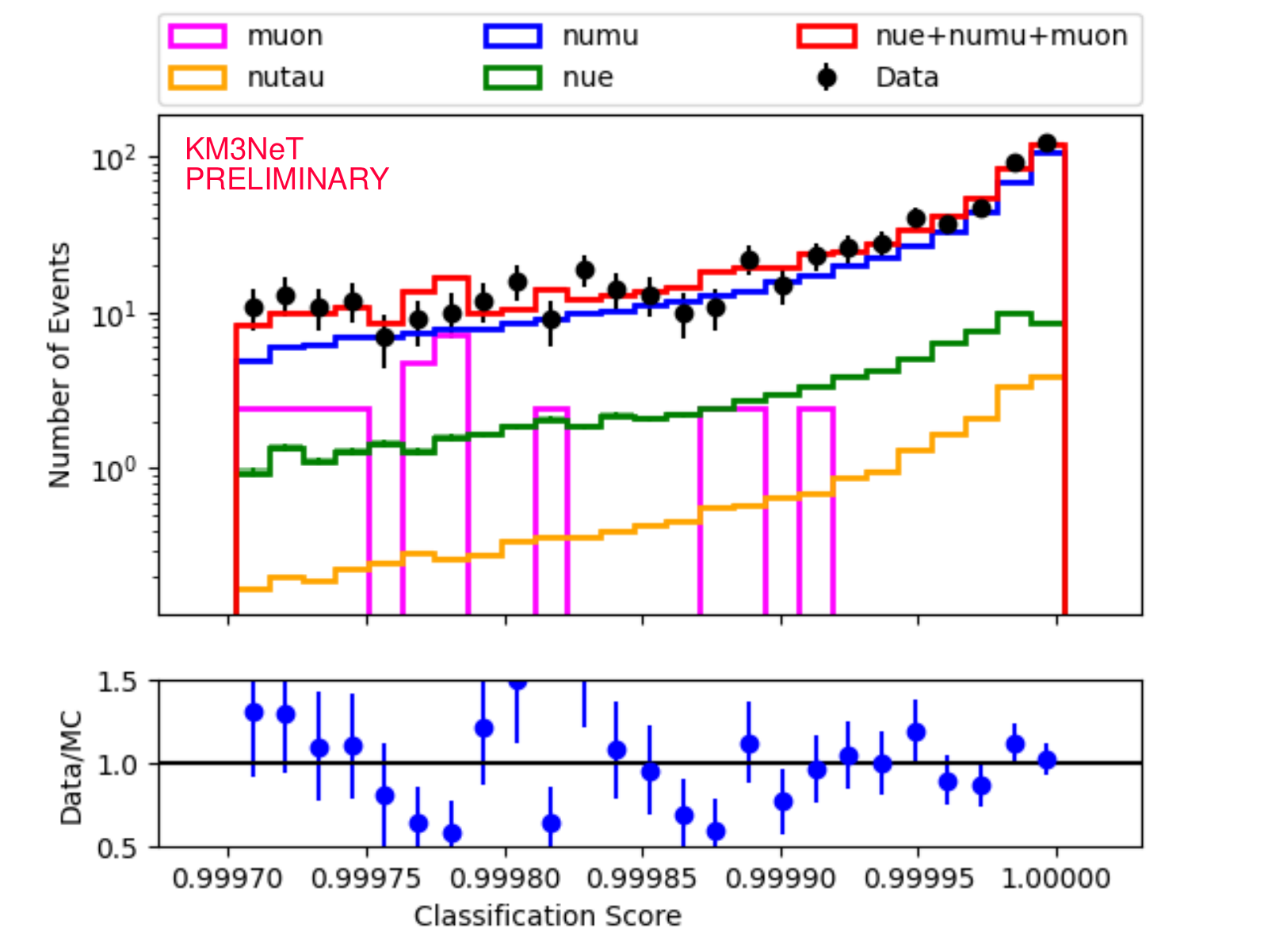}
\includegraphics[width=.46\textwidth,trim=12 0 35 0,clip] {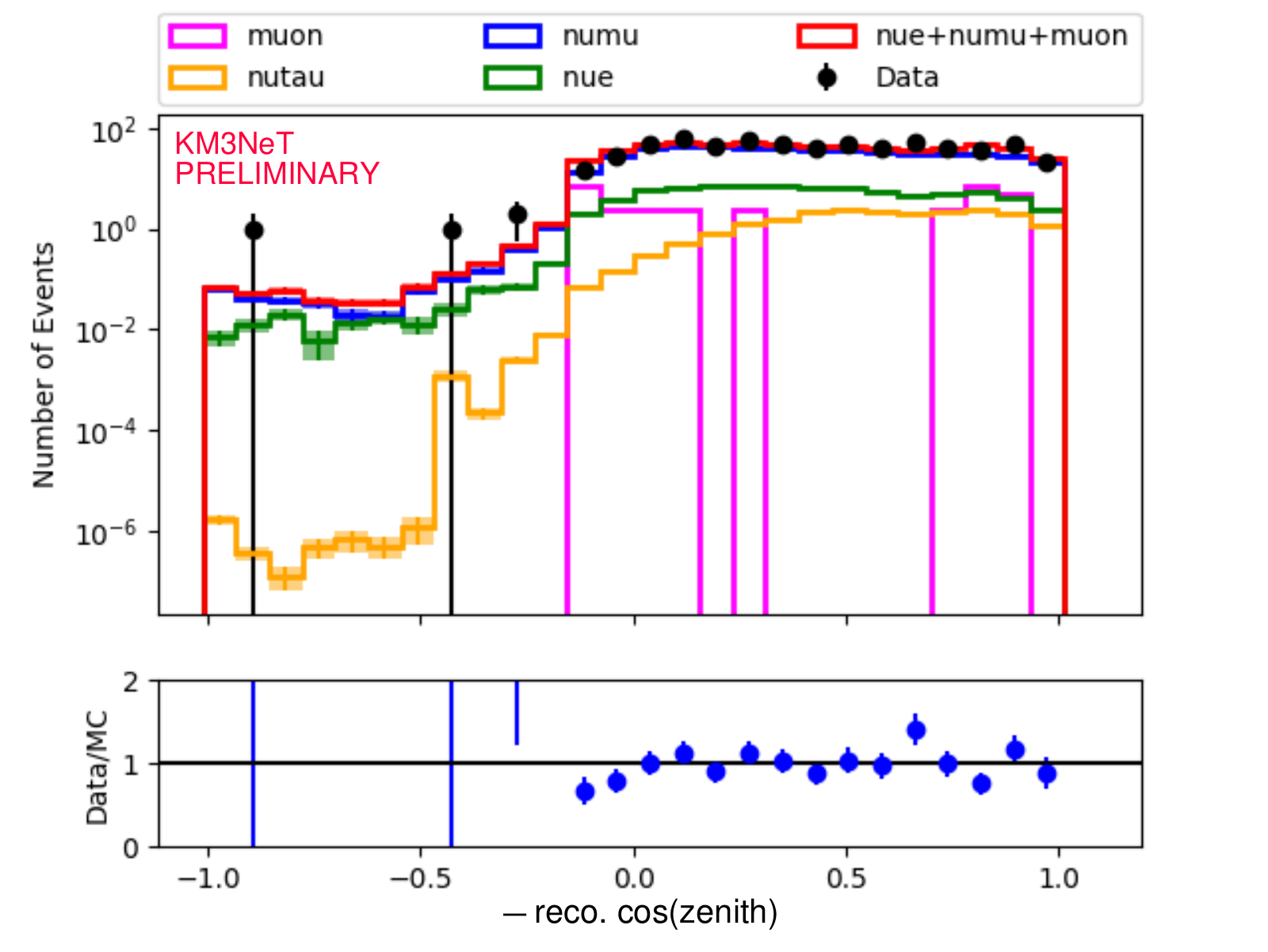}
\qquad
\caption{Data/MC distributions of classification score (left) and - reconstructed cos(zenith) (right).}
\label{fig:4}
\end{figure}

A preliminary test of a selection with an expected 5\% muon is performed on ORCA6 data with 67.8 days of livetime. A good agreement is seen with 631 data events and a MC rate of 631.7 $\pm$ 8.71, among which we have an expected 600.4 $\pm$ 0.60 neutrinos and 31.3 $\pm$ 8.09 muons, where the errors reflect limited MC statistics. Neutrinos are weighted with atmospheric fluxes~\cite{honda} and oscillation parameters with NuFIT5.0~\cite{nufit}; normal hierarchy is used. No detector systematic uncertainties are considered here so far. More data runs will be processed (over a year of data is available). Figure~\ref{fig:4} shows the data/MC comparisons. Optimizations for transient source searches are underway.

\subsection{Alert Distribution}
The GCN alert receiver has been implemented and in operation since November 2020 with a pre-selection of the notice types. It'll be completed with other trigger providers, e.g. TNS\protect\footnotemark  and LSST/ZTF~\cite{LSST, ztf} brokers. For the alert sender, the implementation of Comet broker~\cite{comet} was done, and tests for sending and receiving VOEvents~\cite{voevent} with dummy servers were performed.

\footnotetext{Transient Name Server, https://www.wis-tns.org/}

\section{Summary \& Next Steps}
In summary, the KM3NeT MeV CCSN monitoring pipeline is fully functional and connected to SNEWS. The GeV-PeV neutrino pipeline has a fast online track reconstruction and classification. The alert sender and receiver are mostly ready. A preliminary event selection based on the online classifier has been tested on data and gives good data/MC comparison. Current ongoing work includes online correlation analysis, shower classifier training, and the implementation of the framework for ARCA. Alerts are planned to be sent out beginning in 2022, they will be sent privately during the commissioning phase. After full validation of the alert system, the main KM3NeT alerts will be distributed publicly.



\end{document}